\documentclass[conference]{IEEEtran}
\IEEEoverridecommandlockouts
\usepackage{cite}
\usepackage{amsmath,amssymb,amsfonts}
\usepackage{algorithmic}
\usepackage[final]{graphicx}
\usepackage{textcomp}
\usepackage{xcolor}
\usepackage{adjustbox,multirow}
\def\BibTeX{{\rm B\kern-.05em{\sc i\kern-.025em b}\kern-.08em
    T\kern-.1667em\lower.7ex\hbox{E}\kern-.125emX}}

\usepackage[labelsep=period]{caption}
\renewcommand{\thetable}{\Roman{table}}
\def\tablename{TABLE}

\usepackage{pgfplots}
\usepackage{grffile}
\pgfplotsset{compat=newest}
\usetikzlibrary{plotmarks}
\usetikzlibrary{arrows.meta}
\usepgfplotslibrary{patchplots}

\usepackage[hidelinks]{hyperref}
\usepackage{float}
\usepackage[caption = false]{subfig}

\usepackage{booktabs,array}

\usepackage{mathtools}
\usepackage{siunitx}
\usepackage{bmpsize}
\usepackage{adjustbox}
\usepackage[abs]{overpic}
\usepackage{multirow}

\usepackage{colortbl}%
\newcommand{\myrowcolour}{\rowcolor[gray]{0.925}}

\DeclareRobustCommand*{\IEEEauthorrefmark}[1]{%
	\raisebox{0pt}[0pt][0pt]{\textsuperscript{\footnotesize #1}}%
}

\renewcommand{\baselinestretch}{1.00}
\begin{document}

\title{Organ-Based Chronological Age Estimation\\ Based on 3D MRI Scans}

	\author{\IEEEauthorblockN{Karim Armanious$\,^{\star}$\,\IEEEauthorrefmark{1}\textsuperscript{,}\IEEEauthorrefmark{2},
		Sherif Abdulatif$\,^{\star}$\,\IEEEauthorrefmark{1},
		Anish Rao Bhaktharaguttu\IEEEauthorrefmark{1},
		Thomas K\"ustner\IEEEauthorrefmark{2}\textsuperscript{,}\IEEEauthorrefmark{3},\\
		Tobias Hepp\IEEEauthorrefmark{2},
		Sergios Gatidis\IEEEauthorrefmark{2},
		Bin Yang\IEEEauthorrefmark{1}\thanks{$^{\star}$These authors contributed to this work equally.}}
	\IEEEauthorblockA{\IEEEauthorrefmark{1}University~of~Stuttgart~,~Institute~of~Signal~Processing~and~System~Theory,~Stuttgart,~Germany\\
		\IEEEauthorrefmark{2}University~of~T\"ubingen,~Department~of~Radiology,~T\"ubingen,~Germany\\
		\IEEEauthorrefmark{3}King's~College~London,~Biomedical~Engineering~Department,~London,~England}}

\maketitle

\begin{abstract}
Individuals age differently depending on a multitude of different factors such as lifestyle, medical history and genetics. Often, the global chronological age is not indicative of the true ageing process. An organ-based age estimation would yield a more accurate health state assessment. In this work, we propose a new deep learning architecture for organ-based age estimation based on magnetic resonance images (MRI). The proposed network is a 3D convolutional neural network (CNN) with increased depth and width made possible by the hybrid utilization of inception and fire modules. We apply the proposed framework for the tasks of brain and knee age estimation. Quantitative comparisons against concurrent MR-based regression networks and different 2D and 3D data feeding strategies illustrated the superior performance of the proposed work.
\end{abstract}


\section{Introduction}
\label{sec:intro}

Age estimation is an integral part of biological profile assessment, especially in the medical and forensic domains. In recent years, age estimation has become an important factor for a variety of legal proceedings like immigration, unknown birth date and other criminal/civil disputes \cite{ref1}. There are several methodologies for age estimation based on magnetic resonance (MR) images of the foot, elbow, wrist and teeth \cite{ref2,ref3,hand3D}.

Most commonly, age estimation is conducted on brain MR scans. This is because the human brain undergoes significant morphological changes owing to the ongoing aging process. More specifically, the gray matter (GM) mass is a strong indicator for the biological brain age \cite{ref4}. It was found that the GM mass declines linearly with age \cite{ref5}. In a previous work, T1-weighted MRI scans of the GM mass were utilized to estimate the brain age of the test subject. This could help in the early diagnosis of neurogenerative disorders such as Alzheimer and Parkinson \cite{ref6,ref7}.

Classical approaches for MRI-based brain age estimation involve the manual extraction of features such as cortical thickness and surface curvature \cite{ref8}. This is followed by traditional machine learning algorithms such as kernel methods \cite{ref9} or support vector machines (SVM) \cite{ref10}. A major downside of feature extraction is that the features may not fully represent the contents of the input MR scans. With the advent of deep learning (DL), neural networks greatly increased the accuracy of prediction tasks without manual feature engineering such as in different fields of studies \cite{ref154}. For example, convolutional neural networks (CNNs) were utilized to predict the brain age from 2D input MR slices \cite{ref14,ref144}. More notably, shallow 3D CNN architecture were used to predict the brain age using T1-weighted MRI volumes \cite{ref15,ref16}. 

Within the context of age estimation, it is important to distinguish the difference between the chronological and biological ages. The chronological age (CA) is defined as the amount of time passed since birth to a given date. However, the of biological age (BA) is the actual age of various organs that could be healthier or more damaged owing to different factors such as lifestyle, genetics, medical history and food
habits. Thus, an individual's BA could deviate from his/her CA. Even more, different organs of the same person may have different BAs. This motivates an organ-specific approach to age estimation. Consequently, age estimation of MRI scans could be used to gather important clinical information when observing discrepancies between the predicted BA and the actual CA \cite{ref33}. However, since there is no solid definition of BA, in this preliminary study we consider the CA as the ground-truth label. Therefore, in the following the term age refers to the CA.

In this paper, we propose a new hybrid 3D-CNN architecture for chronological organ-specific age estimation from MRI scans. The proposed network utilizes recent deep learning advances such as Inception and SqueezeNet architectures \cite{ref20,ref21}. We apply the proposed framework on the task of brain CA estimation using an open source brain MR dataset.
Additionally, two data feeding approaches are investigated. The first approach utilizes the entire volume while in the second approach the data is split into more localized 3D chunks. A 2D variant of the proposed network is also presented for computational efficiency purposes. Finally, a quantitative comparison is carried out against current 2D and 3D MR-based age estimation approaches. 

Apart from the brain, other anatomical regions of the human body also display ageing traits such as the joints and knees \cite{ref23}. Thus, to demonstrate the generalization capability of our framework to different body organs, we also apply our framework to MR-based knee age estimation. This could benefit the early diagnosis of knee ailments such as predicting the risk of Arthritis caused by degeneration of the knee cartilage and specifically Osteoarthritis in case of athletes \cite{ref25}.

\begin{figure*}
	\centering
	\includegraphics[width=1\textwidth]{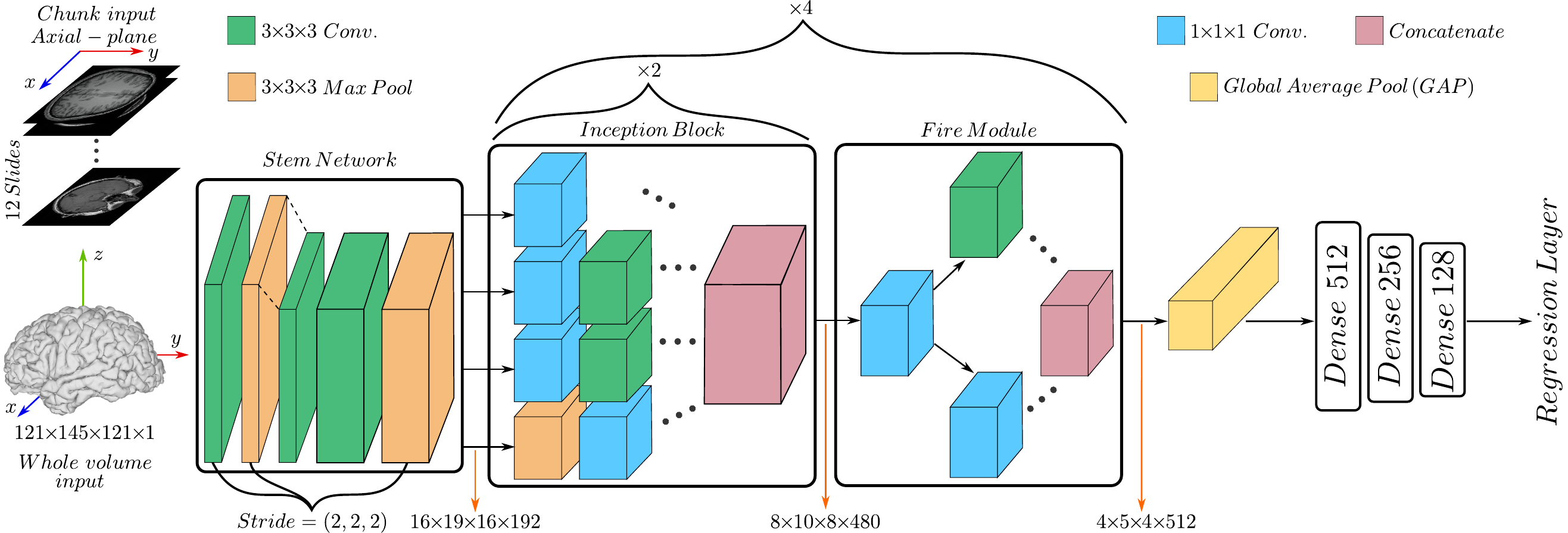}
	\caption{An overview of the proposed 3D hybrid network for MR-based organ age estimation.\label{fig:architecture}}
	\vspace{-3mm}
\end{figure*}

In the following section, the utilized pre-processing steps required for the brain age estimation will be outlined. Accordingly, the proposed architecture and different methods for feeding the input data will be explained. Section~\ref{sec:experiment evaluation} discusses the experimental setup for both the brain and knee datasets. Finally, the results and discussion will be presented in Section~\ref{sec:results and discussions} followed by the conclusion. 

\section{Methods}
\label{sec:methods}

\subsection{Pre-processing of Brain MRI}
The GM is a strong representation of the brain age \cite{ref14}. Thus, we utilize the pre-processing steps recommended in \cite{ref5}. First, the GM tissues are segmented from the input brain volumes using the statistical parameter mapping 12 (SPM12) software \cite{ref18}. The resultant segregated tissues are then spatially normalized according to the international consortium for brain mapping (ICBM) 152 template. Using DARTEL \cite{ref19}, the GM volumes are resampled using a 4 mm smoothing kernel which results in a final volume size $121 \times 145 \times 121$ voxels. Examples of brain MR images in the axial plane are depicted in Fig.~\ref{kneemri}.

\subsection{Architecture}
An outline of the utilized network is shown in Fig.~\ref{fig:architecture}. The proposed 3D-CNN regression is a combination of Inception v1 \cite{ref20} and Squeezenet \cite{ref21} architectures. The inception modules enable learning deeper feature representations by convolving the input volumes with filters of different kernel sizes. This allows the creation of deeper and wider networks while maintaining the same computational budget. Additionally, squeeze and expand layers are also utilized. They form together the fire module which is the prime architectural component of the Squeezenet. These blocks consist primarily of $1 \times 1 \times 1$ kernels and help to downsize the number of parameters even further. 

The final architecture consists of a stem network utilizing $3 \times 3 \times 3$ convolutions and max pooling layers followed by 4 modules, each composed of 2 inception blocks and a single fire module \cite{ref20}. Afterwards, a global average pooling (GAP) is applied which acts as a structural regularizer. This reduces the four-dimensional input tensor to a one-dimensional vector of 512 features which is then connected to three dense layers. Finally, a regression layer outputs the predicted age.

\subsection{3D Input Pipeline}
The manner of feeding the input 3D volumes can have a significant effect on the resultant regression result. In this work, we investigate two different approaches of feeding the input data. The first approach is based on feeding the entire 3D volume as input which results in a single estimated age for each test subject. In order to prevent overfitting, additional data augmentation is applied such as horizontal flipping and  translating the volume within a defined voxel-range. In terms of training efficiency, training on the whole 3D volume demands huge memory space as well as increased training time owing to the on-the-fly data augmentations.

The second approach consists of dividing the MRI volume of each patient into smaller non-overlapping 3D chunks which serve as individual input samples. This implicitly expands the dataset size without further data augmentation. As a result, the network outputs multiple age estimations for each test subject with the final age given as the arithmetic mean of the predicted age of all the input chunks. This could enable a more localized analysis of the predicted age of each input chunk.

For lower computational complexity, a 2D variant (2D-CNN) of the proposed 3D framework is introduced. This network utilizes 2D convolution operations instead of the 3D counterpart while using the same design as in Fig.~\ref{fig:architecture} .

\begin{figure*}
	\subfloat[]{\includegraphics[width = 1.4in, height = 1.4in]{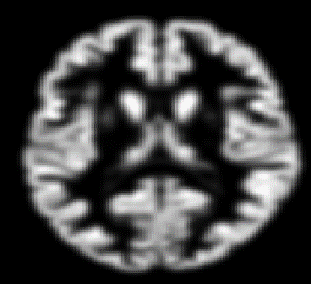}}
	\hfill
	\subfloat[]{\includegraphics[width = 1.4in, height = 1.4in]{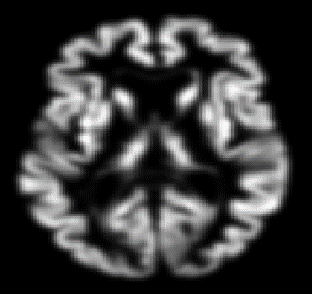}}
	\hfill
	\subfloat[]{\includegraphics[width = 1.4in, height = 1.4in]{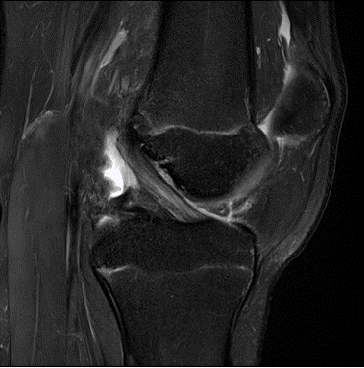}}
	\hfill
	\subfloat[]{\includegraphics[width = 1.4in, height = 1.4in]{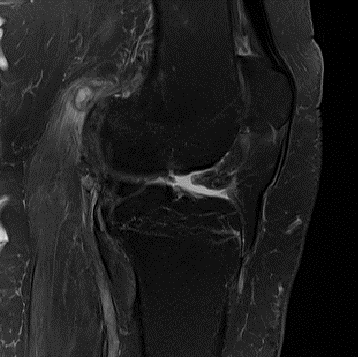}} 
	\caption{Examples for the input MR images. First, segmented GM from axial 3D T1-weighted scans for (a) 23 year old subject, (b) 86 year old subject, and T2-weighted sagittal images for the knee in a (c) 25 year old subject, (d) 88 year old subject.
	}
	\label{kneemri}
	\vspace{-2mm}
\end{figure*}
\begin{figure}
	\centering
	\resizebox{1\columnwidth}{!}{
		\hspace{-1.5mm}
%
%
\definecolor{mycolor1}{rgb}{0.00000,0.44700,0.74100}%
\definecolor{mycolor2}{rgb}{0.85000,0.32500,0.09800}%
\begin{tikzpicture}

\begin{axis}[%
width=4.421in,
height=2.83in,
at={(0.758in,0.481in)},
scale only axis,
xmin=7,
xmax=95,
xlabel style={font=\color{white!15!black}},
xlabel={\Large Age [years]},
yticklabel style = {font=\large},
xticklabel style = {font=\large},
ymin=0,
ymax=50,
ylabel style={font=\color{white!15!black}},
ylabel={\Large \# Patients},
axis background/.style={fill=white},
legend style={at={(0.012,0.825)}, anchor=south west, legend cell align=left, align=left, draw=white!15!black, font=\large}
]
\addplot[ybar interval, fill=mycolor1, fill opacity=0.6, draw=black, area legend] table[row sep=crcr] {%
	x	y\\
	18	5\\
	20.8	17\\
	23.6	35\\
	26.4	33\\
	29.2	33\\
	32	32\\
	34.8	24\\
	37.6	24\\
	40.4	32\\
	43.2	24\\
	46	22\\
	48.8	14\\
	51.6	27\\
	54.4	25\\
	57.2	38\\
	60	41\\
	62.8	31\\
	65.6	30\\
	68.4	27\\
	71.2	22\\
	74	11\\
	76.8	5\\
	79.6	4\\
	82.4	2\\
	85.2	2\\
	88	2\\
};
\addlegendentry{$\text{Brain, }\mu\text{ = 48.63 years}$}

\addplot[ybar interval, fill=mycolor2, fill opacity=0.6, draw=black, area legend] table[row sep=crcr] {%
	x	y\\
	9	9\\
	12.4	10\\
	15.8	20\\
	19.2	7\\
	22.6	14\\
	26	27\\
	29.4	26\\
	32.8	30\\
	36.2	27\\
	39.6	23\\
	43	36\\
	46.4	40\\
	49.8	40\\
	53.2	41\\
	56.6	43\\
	60	37\\
	63.4	23\\
	66.8	35\\
	70.2	16\\
	73.6	16\\
	77	14\\
	80.4	7\\
	83.8	5\\
	87.2	1\\
	90.6	1\\
	94	1\\
};
\addlegendentry{$\text{Knee, }\mu\text{ = 48.89 years}$}
\end{axis}
\end{tikzpicture}%
	}
	\caption{The distribution of training subjects utilized in the brain and knee datasets with $\mu$ representing the mean age.}
	\label{fig:ageDist}
	\vspace{-5mm}
\end{figure}
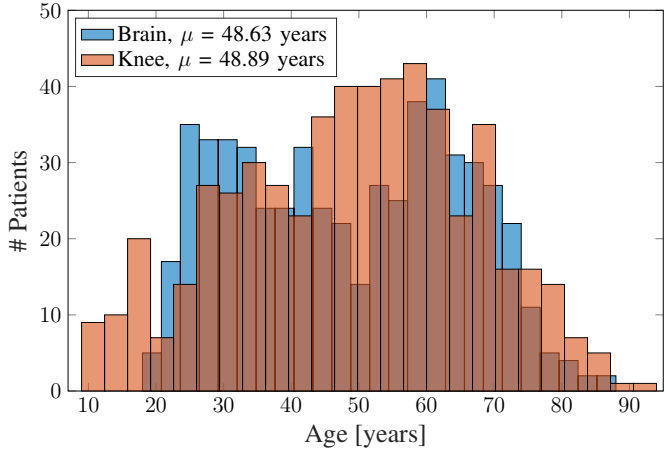
\section{Experimental Evaluations}
\label{sec:experiment evaluation}
\subsection{Brain Dataset}
The proposed 3D CNN architecture is evaluated for the task of brain age estimation using the open source IXI Brain Dataset \cite{ref22}. T1-weighted MR scans from 425, 90 and 45 subjects were used for training, cross-validation and testing, respectively. For the first 3D input pipeline approach (3D-CNN-full), the entire pre-processed GM volumes, each of size $121 \times 145 \times 121$, are fed to the network. For the second approach, individual non-overlapping 3D chunks from each test subject, of size $121 \times 145 \times 12$, are fed instead (3D-CNN-chunk) with a total of approximately 10 chucks for each patient. Additionally, for the 2D variant of the proposed network (2D-CNN), only individual 2D axial slices of $121 \times 145$ pixels are fed into the network. 

\subsection{Knee Dataset}

For the task of knee CA estimation, we utilize the open-source NYU fastMRI dataset \cite{ref29} demonstrated in Fig.~\ref{kneemri}. The dataset consists of T2-weighted scans with fat suppression for 550 subjects. For pre-processing, the voxels of the MR scans were normalized to have zero mean and unit variance with a final volume size of $320 \times 320 \times 28$. The subjects were split into 415, 90 and 45 for training, cross-validation and testing, respectively. To train the 3D-CNN-chunk network, each input volume was split to a chunk size of $320 \times 320 \times 4$ (7 chucks for each patient). Analogously, individual axial slices with a matrix size of $320 \times 320$ were fed as input to the 2D variant of the proposed network. The data distribution of the training dataset for both knee and brain MR are depicted in Fig.~\ref{fig:ageDist}.

\begin{table}[!]
	\caption{Quantitative comparison for CA estimation.\label{tab:results}}
	\centering
	\setlength\arrayrulewidth{0.05pt}
	\small
	\bgroup
	\def\arraystretch{1.15}
	\resizebox{\columnwidth}{!}{%
		\begin{tabular}{l cccc}
			\toprule
			\multirow{2}{*}{Model} & \multicolumn{4}{c}{(a) Brain age estimation}\\\cmidrule(l){2-5} & MAE & SD & Bias & RMSE\\ 
			\midrule  
			\myrowcolour 2D-Huang (\hspace{-0.07mm}\cite{ref14}) & 3.529 & 4.302 & 1.252 & 4.480\\
			3D-Ueda (\hspace{-0.07mm}\cite{ref16}) & 3.705 & 4.298 & 1.267 & 4.481\\
			\myrowcolour 2D-CNN & 3.213 & 4.167 & 1.647 & 4.481\\
			3D-CNN-full & 2.658 & \textbf{3.532} & \textbf{0.590} & \textbf{3.581}\\
			\myrowcolour 3D-CNN-chunk & \textbf{2.283} & 3.546 & 0.904 & 3.659\\
			\bottomrule
			\multirow{2}{*}{Model} & \multicolumn{4}{c}{(b) Knee age estimation}\\\cmidrule(l){2-5} & MAE & SD & Bias & RMSE\\
			\midrule
			\myrowcolour 2D-Huang (\hspace{-0.07mm}\cite{ref14}) & 4.279 & 6.142 & 0.558 & 6.167\\
			3D-Ueda (\hspace{-0.07mm}\cite{ref16}) & 5.354 & 7.056 & \textbf{0.435} & 7.069\\
			\myrowcolour 2D-CNN & 3.721 & 5.351 & 1.407 & 5.533\\
			3D-CNN-full & 3.357 & 4.624 & 0.716 & 4.679\\
			\myrowcolour 3D-CNN-chunk & \textbf{2.996} & \textbf{4.492} & 0.658 & \textbf{4.540}\\
			\bottomrule
		\end{tabular}
	}
	\egroup
	\vspace{-2mm}
\end{table}

\subsection{Experiments}

To demonstrate the performance of the proposed network, we quantitatively compare against other competing approaches for MRI-based brain age estimation. First, the 2D framework from \cite{ref14}, which consists of a modified VGGNet, was compared to the 2D variant of our proposed framework. Additionally, we compare our 3D models (3D-CNN-full and 3D-CNN-chunk) against the recent state-of-the-art 3D network in \cite{ref16}. To ensure a fair comparison, the models were trained using the hyperparameters recommended by the authors. For the quantitative metrics, the mean absolute error (MAE), standard deviation (SD), bias and the root mean square error (RMSE) are utilized.
\vspace{-1mm}
\section{Results and Discussion}
\label{sec:results and discussions}
The quantitative comparisons between MR-based regression networks for both brain and knee CA estimation are presented in Table~\ref{tab:results}. For brain CA estimation, the 2D VGG-based network by Huang \cite{ref14} and the 3D network proposed by Ueda in \cite{ref16} yielded comparable performance with the former resulting in a marginally improved MAE score. The 2D variant of our proposed network outperforms the previous approaches across the utilized metrics. This is due to the increased network depth and width, and thus learning capability, made possible by the utilization of inception and fire modules. Furthermore, the proposed 3D network resulted in the best quantitative scores. However, the manner of feeding the input 3D data had a significant effect on the regression performance. Feeding smaller 3D chunks as inputs via the 3D-chunk network resulted in the best MAE score of 2.283 years (a gain of 0.375 years). However, this came at the cost of a significant deterioration of the bias and, thus, the RMSE.

For the knee CA estimation, similar results can also be observed from Table~\ref{tab:results}. In terms of MAE, the 3D-Ueda model resulted in the worse MAE of 5.354 years compared to 4.279 years by the 2D-Huang approach. We hypothesize that this is the result of reducing the depth of the 3D network, in comparison to the 2D VGGnet utilized 2D-Huang, to accommodate the memory requirements and complex operations necessary for voxel-based training. Furthermore, our proposed networks have significantly better performance with an MAE score of 2.996 years by the chunk data feeding strategy (3D-CNN-chunk) in comparison to 3.357 years for feeding the full MR volumes (3D-CNN-full) with no deterioration in the bias contrary to the results of brain MRI. Based on the above results, we conclude that the chunk-based data feeding strategy is more advantageous than the typical approach of feeding the full MR volumes due to yielding a significantly better MAE metric in both brain and knee age estimations. Additionally, it is clear that utilizing the proposed 3D architecture with sufficient depth enables the utilization of all spatial information and, thus, yields better regression performance.

However, this work is not without limitations. To comprehensively assess the CA estimation performance, we plan to apply the proposed networks on datasets which contain different pathologies such as Alzheimer and Arthritis. Moreover, we plan to further evaluate the proposed networks by conducting experiments on specific age groups and different contrast data.

%
%
%
%
%

\section{Conclusion}
\label{sec:conclusion}
In this work, we present a new network architecture for MR-based organ age estimation. The proposed architecture is based on a combination of inception and fire modules which enables increased network depth and width while maintaining the computational complexity and number of parameters. To showcase the performance of the presented framework, quantitative comparisons are carried out against current state-of-the-art MR regression networks for the tasks of brain and knee age estimation. The proposed architecture resulted in a significantly improved performance.

In the future, we would like to explore the tasks of age estimation for additional organs which may exhibit age-related changes such as spine, kidneys and liver. Additionally, we plan to conduct clinical analysis by radiologists for the cases were the estimated age deviates remarkably from the ground-truth chronological age in order to infer the BA.

%

\bibliographystyle{IEEEtran}

\end{document}